\documentstyle[seceq,mbf,wrapft,amsbsy,amssymb]{ptptex}

\title{ 
Gravity beyond linear perturbations in the braneworld
}

\author{ 
    Hideaki {\sc Kudoh}\footnote{E-mail: kudoh@yukawa.kyoto-u.ac.jp}
}

\inst{ 
Department of Physics, Kyoto University, Kyoto 606-8502, Japan
}
  
\abst{ 
In the present paper, we briefly review recent studies of second-order gravitational perturbations in braneworld models. After we consider the possibility of pathological behavior of gravity at higher orders of perturbation, second-order perturbations are discussed in Randall-Sundrum braneworld models. Because the mass spectra in these braneworld models are different, we analyze them using different approaches.  In the respective models, it is discussed that 4D Einstein gravity is approximately recovered at level of second-order perturbation, although there are some exceptional cases in the two-brane model.
\\
\\
KUNS-1787
}
 
\begin{document}

\maketitle

\section{Introduction}

The idea of the braneworld has attracted considerable attention from many points of view.\cite{Arkani-Hamed:1998rs,Randall:1999ee,Randall:1999vf}  In particular, the models introduced by Randall and Sundrum (RS) \cite{Randall:1999ee,Randall:1999vf} have been studied by many people because the so-called warped compactification of RS models make the models intriguing.

A significant point in discussing the viability of these models is whether 4D Einstein gravity is recovered on the brane from 5D Einstein gravity in the bulk. In the RS models, the gravity that is induced on a flat brane by matter fields confined there has the form \cite{Shiromizu:2000wj}
\begin{eqnarray}
{}^{(4)}G_{\mu\nu}
=  8 \pi G\,  T_{\mu\nu}+ (8 \pi G_5)^2 \pi_{\mu\nu}
-E_{\mu\nu} , 
\label{eq:effective 4D}
\end{eqnarray}
where $8\pi G$ and $8\pi G_5 ( := \kappa)$ are the 4D and 5D effective gravitational coupling constants, respectively, and $\pi_{\mu\nu}$ is a tensor quadratic in the 4D energy-momentum tensor $T_{\mu\nu}$.
The projected Weyl tensor $E_{\mu\nu}$ is determined by solving the 5D equations.
Apparently, the last two terms are contributions from the existence of the 5D bulk spacetime. However, the evaluation of $E_{\mu\nu}$ in the above formulation is not easy in general, and explicit and typical contributions in general situations have not been known, except in the case of linear perturbations. \cite{Sasaki:2000mi} 
One fundamental but important fact owing to the warped compactification is that in the RS one-brane model, linearized 4D Einstein gravity is derived on the brane from 5D Einstein gravity when we consider non-relativistic matter on the brane.\cite{Garriga:2000yh,Tanaka:2000er,Mukohyama:2001ks,Mukohyama:2001jv} Furthermore, the conventional FRW universe is realized on the brane at low energies.
These facts imply that the last two terms in Eq. (\ref{eq:effective 4D}) can be omitted in the restricted configurations.

The recovery of linearized 4D Einstein gravity in the RS one-brane model is expected, and it can be understood by considering the equation for linearized gravitational perturbations, which has a volcano potential due to the tension of the brane.\cite{Randall:1999vf}  The volcano potential admits a normalizable bound state, which is called a ``zero mode" because the mass eigenvalue of the 4D d'Alembertian is zero, and also continuum modes that correspond to massive Kaluza-Klein modes (KK).  However, the contribution of massive modes to the induced gravity is small compared with that of the zero mode, owing to the potential barrier, and therefore the zero mode gives the long-ranged force of conventional Newton's law.

For gravity beyond linear perturbations, the above argument does not apply, and for this reason the recovery of 4D linearized Einstein gravity is not expected to be realized in a simple treatment.
In the one-brane model, if we naively write down the 4D effective action, integrating over the extra dimension with the aid of the decomposition with respect to the mass eigenvalues, it seems that the KK mode coupling is ill-defined and diverge.\cite{Tanaka:2000zv}
However, this pathological behavior might be an artifact of the fact that this argument is based on a mode-by-mode analysis using the decomposition of the mass eigenmodes. In fact, the mode-by-mode analysis does not give a good description of the behavior of massive modes.\cite{Tanaka:2000zv,Giddings:2000mu} In the two-brane model, other pathological behavior seems to appear on the negative tension brane; some hierarchically enhanced terms appear at second order in the perturbation, and then seem to prohibit the recovery of the conventional 4D gravity in the nonlinear regime.

Nevertheless, there is the indication that the induced gravity is described well by 4D Einstein gravity, even in the highly nonlinear and non-perturbative regimes, \cite{Wiseman:2001xt,Giannakis:2001zx,Kudoh:2001wb,Kudoh:2001kz,Kudoh:2002mn} and this supports the viability of braneworld models.

In this paper, we review the perturbative approach applied to understanding gravity beyond linear-order perturbations, i.e., second-order perturbations.\cite{Giannakis:2001zx,Kudoh:2001wb,Kudoh:2001kz,Kudoh:2002mn}
In the analyses of second-order perturbations in the RS models, static and axisymmetric configurations, in which the metric on the branes is spherically symmetric, have been assumed for simplicity. In
\S\ref{sec:Single brane case} we review the second-order perturbations in the RS one-brane model without a bulk scalar field. In \S\ref{sec:Two branes case} we study the second-order perturbations in the RS two-brane model where a stabilization mechanism is taken into account. Section \ref{sec:summary} is devoted to a summary.

\section{one-brane model}
\label{sec:Single brane case}

\subsection{Leading-order perturbations}

In this section we review second-order perturbations in the RS one-brane model.\cite{Kudoh:2001wb} The system of this model is simpler than that of the two-brane model, and therefore it is more tractable in the study of gravitational perturbations.

The most general axial-symmetric and static metric has the following form: 
\begin{eqnarray}
    ds^2 =a^2(y)
    \left[ -e^{A(r,y)} dt^2 + e^{B(r,y)} dr^2+ e^{C(r,y)} r^2 
     d\Omega^2 \right] + dy^2 \,.
\label{eq:metric single} 
\end{eqnarray}
Here $a(y)$ is a warp factor and it is taken to be $a=e^{-|y|/\ell}$,  since we consider gravitational perturbations around the RS background solution of  AdS$_5$, in which the vacuum brane of positive tension is located at $y=y_+ \equiv 0$. Non-relativistic matter fields confined on the brane are assumed to have an energy-momentum tensor of the perfect fluid form   
\begin{eqnarray}
    T_{+ \nu }^{ ~~ \mu} = 
    {\mathrm{diag}} 
    \{- \rho_{+}(r), P_{+}(r), P_{+}(r), P_{+}(r) \}.
\label{eq:EM tensor}
\end{eqnarray}

To obtain perturbative equations for the metric functions, we expand $A$, $B$, and $C$ to second order as $A=\sum_{J=1,2} A^{(J)}$, etc.  The 5D Einstein equations with the transverse-traceless (TT) conditions
\begin{eqnarray}
   A^{(1)}  = -\frac{1}{r^2}\partial_r (r^3 B^{(1)} ) ,
\quad
   C^{(1)} =  \frac{1}{2r }\partial_r (r^2 B^{(1)} ) 
\label{eq:TT cond}
\end{eqnarray}
give the master equation for the second order perturbations
\begin{eqnarray} 
\left[
{1\over a^2}\partial_y a^4\partial_y + {\Delta} 
\right]
A^{(J)}  
=   - \epsilon^{(J)} \left[
        \frac{2}{\ell} \int^{y}_{\infty} a^2 Q_{yy} dy 
    +   \frac{1}{r^2} \partial_r \left( r^2 A^{(1)}_{,r}{B}^{(1)} \right) 
    \right],
\label{eq:master eq. A}
\end{eqnarray} 
where $\Delta := \sum_{i=1}^3 \partial_i^2$ and 
$Q_{yy}=\frac{1}{4} \left[ (A_{,y})^2+(B_{,y})^2+2(C _{,y})^2 \right]$. Moreover, we have introduced the symbol $\epsilon^{(J)}$, defined by $\epsilon^{(1)}=0$ and $\epsilon^{(2)}=1$, to represent the first- and second-order equations in a single expression.  
The source terms on the right-hand side are absent at linear order, because they are quadratic in linear-order quantities. 
The other metric functions at second order, $B^{(2)}$ and $C^{(2)}$, are explicitly related to $A^{(2)}$ by constraint equations that are obtained from the 5D Einstein equations.

To solve the master equation, we need to specify the boundary conditions. 
In the original coordinates (\ref{eq:metric single}), the surface of constant $y$ has been chosen so that the metric functions vanish in the limit $y\to \infty$, to coincide asymptotically with AdS$_5$.
After fixing the surface of constant $y$ at infinity, the coordinates are extended to the region near the brane. 
Therefore, the location of the brane does not generally coincide with the $y=y_+$ surface.\cite{Garriga:2000yh} In such coordinates, the junction condition is not trivial. Thus it is convenient to introduce Gaussian normal coordinates $\bar{x}^a$, in which the location of the brane remains at $\bar y=0$,  but the metric form is kept diagonal. 
The boundary conditions in the original coordinates are obtained through gauge transformations defined by $\bar{x}^a=x^a + \xi^a$. 
Israel's junction condition gives the boundary conditions as 
\begin{eqnarray}
  \kappa ^{-1} {A}^{(J)}_{,y}|_{y=0} 
  = \frac{2}{3}\rho_+^{(J)}  
+\epsilon ^{(J)} \Bigl[ P_+ + \kappa ^{-1} (\delta A_{,y}^{(2)})|_{y=0}  \Bigr]\,,
\label{sigma}
\end{eqnarray}
where $\delta A^{(2)}$ is defined as the difference between the metric in the original coordinates (\ref{eq:metric single}) and that in the Gaussian normal coordinates $\bar{x}^a$: $\delta A^{(2)} := A^{(2)} - \bar A^{(2)}$.

The formal solution to the master equation (\ref{eq:master eq. A}) with the boundary conditions is obtained by means of Green's function. We have
\begin{eqnarray}
  a ^2 {{ {A}}}^{(J)}(r,y)  &=& 
  2  \int {dx'}^3   G_A |_{y'=0} ~
   \Bigl[
    \frac{2 \kappa }{3}\rho^{(J)} 
+\epsilon ^{(J)} \Bigl( \kappa P +  \delta A_{,y'}^{(2)}|_{y'=0}  \Bigr)
 \Bigr]  
\cr
  && - 2 \, \epsilon^{(J)} \int d^3x' dy'\, 
          G_A 
    \left[ \frac{  ( r^2 A_{,r} B  )_{,r} }{r^2} 
    + \frac{2}{ \ell}  
        \int_{\infty}^{y'} a^{2}  Q_{yy}  dy'' \right] , 
\quad
\label{eq:formal sol. of sA^J}
\end{eqnarray}
and Green's function in the static case is given by 
\begin{eqnarray}
    G_A({\boldsymbol x},y;{\boldsymbol x}',y') 
= - 
\int  \frac{d^3 {\boldsymbol k} e^{i{\boldsymbol{k}}  ({\boldsymbol x}-{\boldsymbol x'})} }{(2\pi)^3}  
    \left[
        \frac{N a(y)^2 a(y')^2 }{ {\boldsymbol k}^2+\epsilon^2}
     +  \int dm  \frac{u_m(y) u_m(y')}{m ^2+ {\boldsymbol k}^2 }  
   \right], \qquad
\label{eq:static Green fun.}    
\end{eqnarray} 
where the $u_m(y)$ are mode functions, and $m$ is the 4D mass eigenvalue. The normalization factor $N$ is defined by 
\begin{eqnarray}
\frac{1}{N} := 2\int_{y_{+}}^{y_{-}} a^2 dy,
\end{eqnarray}
where $y_-$ is set to infinity in the present model. 
The orthogonality of the mode functions is expressed by 
\begin{eqnarray}
 \int^{\infty}_{0} {dy} \  u_m  =0  \quad(m\ne 0), 
\quad
  2 \int _0 ^\infty \frac{dy}{a^{2}}  u_m  u_{m'} = \delta (m-m') \, .
 \label{eq:orthogonality of u_m u_m}
\end{eqnarray}
The mode decomposition and its meaning are obvious from Green's function. The first term in Eq.~(\ref{eq:static Green fun.}) is the contribution from the zero mode ($m=0$), and the second term corresponds to the propagator due to KK states that have non-zero mass eigenvalues ($m>0$).

The leading-order in linear perturbations is obtained from the zero-mode truncation. Substituting only the zero-mode part into Eq. (\ref{eq:formal sol. of sA^J}) and transforming it into Gaussian normal coordinates, we obtain the leading-order in linear perturbations induced on the brane:
\begin{eqnarray}
- \bar A^{(1)}_{0}
= \bar B_{0 }^{(1)}= \bar C_{0 }^{(1)} \approx  -2 \Phi_+ .
\quad 
({\mathrm at~} \bar y=y_+)
\label{eq: A0=4D ein}
\end{eqnarray}
Here we have represented the result by using the Newton potential, which is defined by
\begin{eqnarray}
    \Delta \Phi_+ (r)
    &:=&
      4 \pi G ~ \rho^{(1)}_{+}(r)   \,,
\label{eq:def Newton potential}
\end{eqnarray}
where the 4D Newton's constant is defined by $8\pi G := \kappa N$.
Hereafter, the gauge freedom of the radial coordinate is chosen to be the isotropic gauge.
Now, the evaluation of leading-order terms in second-order perturbations is straightforward. 
Substituting the zero-mode part of Green's function and the first-order quantities, we finally obtain, at the leading order, 
\begin{eqnarray}
 \Delta  {\bar A }^{(2)} (r,0)
&\approx& 
 8\pi G   \left(\rho_+^{(2)}+ 3P_+^{(2)}  -2\Phi_+ \Delta \Phi_+ \right), 
\cr
\Delta \bar B^{(2)} (r,0)
= 
\Delta {\bar{ {C}}^{(2)} } (r,0)
&\approx&
  - 8 \pi G \rho^{(2)}_{+} 
  + 4 \Phi_+ \Delta \Phi_+   
  - \left( \Phi_{+,r} \right)^2
.
\label{eq:leading for A2 B2}
\end{eqnarray}
These results agree with those for 4D Einstein gravity.

\subsection{Suppression of the KK mode propagation}

The appearance of conventional 4D gravity can be easily understood using the zero-mode truncation of Green's function. Thus our main task is to demonstrate the suppression of contributions due to KK modes. To do this, we need to evaluate a convolution of Green's functions. Because $G_A$ is composed of a zero mode part $G_0$ and a KK mode part $G_K$, the contribution from these terms can be decomposed into several pieces, depending on which combination of the three propagators is used.

We note that the mode couplings that contain two zero modes, such as $G_K \times G_0 \times G_0$, vanish because of the orthogonality of mode functions. 
Thus our concern is the mode couplings that include at least two KK modes.
Since the mode-by-mode analysis yields pathological behavior even at the level of linear perturbations, it is necessary to sum up all the mass eigenvalues to handle the KK mode interactions.\cite{Tanaka:2000zv}
Carefully performing such analysis,\cite{Kudoh:2001wb} it is found that the leading order of the corrections is 
\begin{equation}
  O\left( \frac{\Phi_+^2}{r_\star^2} \right)  O\left( \frac{\ell^2}{ r_\star^2} \ln \left( \frac{r_\star}{\ell} \right) \right),
\end{equation}
where $r_\star$ is a typical length scale of perturbations. 
Therefore, the corrections due to the KK modes are suppressed on scales sufficiently large compared to the 5D curvature scale $\ell$. 
Note that this suppression essentially arises from the following inequality involving Green's function:
\begin{eqnarray}
 &0 \le -G_A ( {\boldsymbol x}_1,y; {\boldsymbol x}_2',0) \lesssim 
 \frac{a^2(y)}{4\pi\ell\,R_{12}^2 }
 ( \ell + R_{12} ), &
\cr 
& \int_0^{\infty} \frac{dy}{ a^{ 2}}  G_K({\boldsymbol x}_1,y;{\boldsymbol x}_2,0)
   G_K({\boldsymbol x}_3,y;{\boldsymbol x}_4,0)
  < \frac{1}{ 4(4\pi)^2 R_{12} R_{34} (R_{12}+R_{34})(\ell +R_{12}+R_{34})},
&
 \nonumber
\end{eqnarray}
where $R_{AB}\equiv |{\boldsymbol x}_A-{\boldsymbol x}_B|$.

\section{Two-brane model}
\label{sec:Two branes case}

\subsection{Master equations}

In this section, we consider second-order perturbations in the RS two-brane model.\cite{Kudoh:2001kz,Kudoh:2002mn} The second brane, which has negative tension, is located at $y=y_- ~(> y_+)$. Hence, there is a new characteristic length scale of this system, the distance between the two branes, which we refer to as the ``radius". Because this radius is related to the hierarchy,\cite{Randall:1999ee} it must be stabilized at an appropriate value.  
As a radius stabilization mechanism, we introduce a bulk scalar field, following Goldberger and Wise.\cite{Goldberger:1999uk}
The Lagrangian for the scalar field is 
\begin{equation}
 {\cal  L}= - {1\over 2} \tilde  g^{ab} 
     \tilde\varphi_{,a} \tilde\varphi_{,b}
          -V_B(\tilde\varphi)
          -\sum_{\sigma=\pm} V_{(\sigma)}(\tilde\varphi)
              \delta(y-y_{\sigma}) \,.
\label{eq:Lagrangian}
\end{equation}
The scalar field is expanded up to second-order as 
$ \tilde\varphi = \phi_0(y) + \varphi^{(1)}(r,y)+\varphi^{(2)}(r,y)$, where $\phi_0$ is the background scalar field configuration.
The gravitational perturbations, however, cause the radius to fluctuate, and we cannot assume the metric (\ref{eq:metric single}) in the bulk. Instead, it is useful to assume the ``Newton gauge," 
\begin{eqnarray}
 ds^2 
      &=& e^{2Y } dy^2 + a^2 
\left[- e^{A - \psi } dt^2
      + e^{B - \psi } dr^2
      + e^{C - \psi } r^2 d\Omega^2 
\right].
\label{eq:metric}
\end{eqnarray}
Here $A$, $B$, and $C$ correspond to the TT part (\ref{eq:TT cond}), and $\psi$ to the trace part.
The energy-momentum tensors are given in the perfect fluid form as in Eq. (\ref{eq:EM tensor}):
\begin{eqnarray}
    T_{{\pm} \nu }^{ ~~ \mu} = 
    a_\pm^{-4} \,
    {\mathrm{diag}} 
    \{- \rho_{\pm}, P_{\pm}, P_{\pm}, P_{\pm} \}.
\label{eq:EM tensor+-}
\end{eqnarray}
The warp factor in the definition (\ref{eq:EM tensor+-}) is incorporated for the following reason. In the present analysis, we adopt the normalization in which all physical quantities are always measured on the positive tension brane at $y=y_+$. Because a length scale is warped by the warp factor, physical quantities on the negative tension brane at $y=y_-$, such as $\rho_-$ and $P_-$, are measured at $y=y_+$ as $a_-^{-4}\rho_-$ and $a_-^{-4}P_-$. 
Another normalization, which uses the proper length to measure physical quantities, is explained in \S\ref{sec:convention}.  Each normalization has its own virtues.

The warp factor experiences a back reaction from the bulk scalar field, and in general we cannot assume that the warp factor has a pure AdS form.  It is determined by solving the background equation, 
\begin{eqnarray}
    H^2(y) \equiv  \left( \frac{\dot a}{a } \right)^2
    &=& {\kappa \over 6}
    \Bigl( {1\over 2} \dot\phi_0^2(y) 
        - V_B(\phi_0(y))-\kappa^{-1}\Lambda  
    \Bigr).
\label{eq:background eqs}
\end{eqnarray}

As compared with the one-brane model (\ref{eq:metric single}), we have two additional scalar functions in the present case,  $Y$ and $\psi$. They are not independent, being related to each other by the 5D Einstein equations. Moreover, the perturbation of the scalar field $\tilde\varphi$ is also related to $Y$ and $\psi$.  The constraint equations are given as 
\begin{eqnarray}
   && \psi^{(J)} (r,y) = Y^{(J)} + \epsilon^{(J)} \Delta^{-1} S_\psi  \,,
\nonumber
\label{eq: phi^J=Y^J+...}
\\
   &&   \varphi^{(J)} (r,y)
    = \frac{3}{2\kappa \dot \phi_0 a^2}  \partial_y( a^2 Y^{(J)})
    + \frac{3}{2\kappa \dot \phi_0} \epsilon^{(J)} \Bigl[
      S_\varphi 
    + \partial_y \Delta^{-1} 
  S_\psi 
    \Bigr] \,.
\label{eq:d varphi = Y^J +...}
\end{eqnarray} 
Here $S_{\psi}$ and $S_{\varphi}$ are second-order source terms, which are constructed from the first-order quantities.\cite{Kudoh:2001kz} 
Here and hereafter we use second-order source terms, such as $S_*$ and $\mathbb S_*$, without giving their definitions. All of them are defined in Ref. \citen{Kudoh:2001kz}.

The equations for $A^{(J)}$ and $Y^{(J)}$ are obtained from the 5D Einstein equations. The equation for $A$ is basically the same as Eq. (\ref{eq:master eq. A}), except for the source terms on the right-hand side.  The master equation for scalar perturbations is given by 
\begin{eqnarray}
   &&  \left[  a^2\dot\phi_0^2\partial_y{1\over a^{2}\dot\phi_0^2}
      \partial_y a^2-{2\kappa\over 3}a^2\dot\phi_0^2  + \Delta \right] Y^{(J)}= \epsilon^{(J)} S_Y  \,.
\label{eq:Ein-Y^J}
\end{eqnarray}
Let us consider the junction conditions for the scalar field.
Integrating the equation of motion for the scalar field across the brane, we obtain the junction condition in the Newton gauge as 
\begin{eqnarray}
 \frac{2}{\lambda_{\pm}}( \varphi^{(J)} - \dot\phi_0   
 \stackrel{(J)~~}{\hat\xi_{\pm}^y)}
 &=& 
  \mp \frac{3}{ \kappa a^2\dot\phi_0}
     \Delta Y^{(J)}
   + \epsilon^{(J)}  S^{ \pm }_{\rm jun},  
\label{eq:Jun of varphi}
\end{eqnarray}
where we have defined
\begin{equation}
   \lambda_{\pm}
   :={2\over V''_{(\pm)}\mp 2(\ddot\phi_0/\dot\phi_0)} .  
\end{equation}
Through this boundary condition, the potentials $V_{(\pm)}$ on the branes affect the perturbations.
Combining the junction condition (\ref{eq:Jun of varphi}) and the constraint (\ref{eq:d varphi = Y^J +...}), we obtain the boundary condition for (\ref{eq:Ein-Y^J}), and in the end we can construct the formal solution using Green's function.

\subsection{ Derivative expansion }

Formal solutions of the TT part are given by using Green's function (\ref{eq:static Green fun.}), where the continuous mass spectrum in the one-brane model is replaced with the discrete one due to the $S^1/Z_2$ compactification. Because of this difference, we can use another method to evaluate contributions from the discrete massive modes.
The long-ranged contribution of the TT part, $A_0$, is evaluated using the zero-mode truncation, as before. The remaining part, $A_S$, which comes from massive KK modes, is evaluated using a derivative expansion method. We expand the perturbation variables in terms of the expansion parameter $(H r_\star)^{-1}$, assuming that the typical length scale $r_\star$ of perturbations is much longer than the 5D curvature scale $H^{-1}$. 
It is important to stress that this derivative expansion method is valid only when the mass of the first excited mode is sufficiently large. In the one-brane limit ($y_- \to \infty$), the mass spectrum becomes continuous and the mass of the first excited mode approaches zero $(m \to 0)$.  Therefore the derivative expansion method is no longer valid in this limit.

For a scalar-type perturbation, there is no zero-mode, owing to the stabilization mechanism. \cite{Tanaka:2000er}
To study the contributions from massive modes, we expand the perturbation variables in a derivative expansion. 
The scalar-type perturbations are divided into three parts: a pseudo-long-range part $Y_{\mathrm{pse}}$, a short-range part $Y_S$, and a part involving interaction terms $Y_\lambda$.  
The first one consists of the formal solution of the lowest order in the derivative expansion including long-range metric perturbations. 
The short-range part consists of the formal solution at the next order, which is obtained through iteration of the derivative expansion. 
The part involving interaction terms, $Y_\lambda$, is defined to consist of the terms that include a potential on the brane, represented by $\lambda_\pm$ [see Eq. (\ref{eq:Jun of varphi})]. They are explicitly given as 
\begin{eqnarray}
 Y^{(J)} &=& Y^{(J)}_{ {\mathrm{pse}} } + Y^{(J)}_{\lambda} + Y^{(J)}_{S} , 
\nonumber
\\
 \Delta Y^{(J)}_{\mathrm{pse} } (r, y_{\pm})
   &=& 
  - \frac{\kappa N}{3} \sum_{\sigma=\pm} a_{\sigma}^4 
    T^{(J)}_{\sigma}  
    \mp \frac{\kappa H}{3} a_{\pm}^2 T_{\pm}^{(J)}
 - \epsilon ^{(J)} 
 \Bigl(  S_\psi + 2H a_{\pm}^2 S_{\xi}^{\pm}
 \Bigr) 
\cr
&&
   - 2N \epsilon^{(J)} \biggl[
     \sum_{\sigma=\pm} \sigma a_{\sigma}^4
     S_{\xi}^{\sigma} 
   + \int^{y_{-}}_{y_{+}} dy 
 \Bigl(
      a^2 v_{\pm} \Delta {\mathbb{ S}}_\varphi 
    + \frac{3u_{\pm}} {2\kappa \dot \phi_0^2} \Delta {\mathbb S}_Y 
    -a^2 S_\psi 
  \Bigr)
\biggr]  ,
\cr
\Delta  Y^{(J)}_{ \lambda } (r, y_\pm)
&=&
 - 2N  \Delta  \left(  
    \sum_{\sigma= \pm} \sigma  \frac{ L_\sigma^{(J)} }{H_\sigma}
  + \frac{ L_\pm^{(J)} }{a^2_\pm N} 
  \right)  ,
\cr
    Y^{(J)}_{ S} (r, y_{\pm})
 &=&  \frac{3N}{\kappa}
       \int_{y_{+}}^{y_{-}}  
      \frac{u_{ \pm } \Delta Y ^{(J)}_{ 0} }
      { \dot\phi_0^2}   dy'  ,
 \label{eq:Y^J_[GGS] y_+-}
\end{eqnarray}
where  
\begin{eqnarray}
 L_\sigma^{(J)}(r) = H(y_\sigma)
   \biggl[ 
    - \sigma \frac{3 \lambda_{\sigma}}{2\kappa\dot\phi_0^2} \Delta  Y^{(J)}
   + \epsilon^{(J)} a_{\sigma}^2
  \Bigl( 
    \frac{ \lambda_{\sigma}}{2\dot\phi_0} S^{\sigma}_{\rm jun} 
  - \frac{ \ddot\phi_0  \varphi^2}{2 \dot\phi_0^3} 
  + \frac{3 \varphi  \Delta  Y }
      {2\kappa a^2 \dot\phi_0^3}
  \Bigr) 
   \biggr] _{y=y_{\sigma}}\,,
\nonumber
\end{eqnarray}
and the functions $u_\pm$ and $v_\pm$ are given by 
$ u_{\pm} :=1-2 H v_{\pm}$ and 
$ v_{\pm} := a^{-2} \int_{y_{\mp}}^y a^2 dy'$.

These formal solutions for the TT part and the scalar-type perturbations enable us to evaluate the perturbations induced on each brane.

To obtain the induced gravity, we must transform the perturbation quantities calculated in the Newton gauge into those in Gaussian normal coordinates of the form
\begin{eqnarray}
ds^2 = d \bar y^2 + a^2(\bar y) 
\left(  - e^{\bar A} d\bar t^2 +e^{\bar B} d\bar r^2 + e^{\bar C} \bar r^2 d\bar \Omega^2  
\right).
\end{eqnarray}
  We need two sets of Gaussian normal coordinates, that in which the positive tension brane is located at $\bar y= \bar y_+$ and that in which the negative tension brane is located at $\bar y=\bar y_-$. After gauge transformations for each coordinate set, we obtain the metric perturbations induced on the branes,
\begin{eqnarray}
  \Delta \bar A^{(J)} _{0 }(r,y_{\pm})
   &=&  8 \pi G  (\rho ^{(J)}_{\pm} + 3P ^{(J)}_{\pm}  )
 - {\Delta} 
   \Bigl[
     {\hat\xi_{ \pm }^y} \bar{A} _{\pm,y} 
   + {\hat\xi_{ \pm }^r} \bar{A} _{\pm,r}
   + \dot H ( {\hat\xi_{ \pm }^y})^2 
    \Bigr]\nonumber
\label{eq:bar A_0}
\\
 &&  + 2N \sum_\sigma \sigma a_{\sigma}^4 (S_{\xi}^{\sigma} - S_{\Sigma}^{\sigma})
\nonumber
\cr
& &  
 +  {2N}  \int^{y_{-}}_{y_{+}} dy 
  \biggl[ 
    a^2 ( S_{A\pm} - S_{\psi\pm})
  + a^2 v_{\pm} \Delta {\mathbb S}_{\varphi\pm} 
  + \frac{3u_{\pm}}{2 \kappa \dot \phi_0^2}  
    \Delta {\mathbb S}_{Y\pm}
  \biggr] 
  - \Delta Y^{(J)}_{\lambda}, 
\nonumber
\\
&&\hspace{-2cm}   \Delta \bar B^{(J)}  (r,y_\pm )\nonumber\\
  &&\hspace{-1cm} =  - \frac{\Delta \bar A^{(J)}}{2}   
 \mp \frac{\kappa H}{2}a^2_\pm T^{(J)}_\pm
 - \frac{3}{2} \Delta Y^{(J)}
 - \frac{3}{2} \epsilon^{(J)}
\left(
    \frac{2}{3}\Delta  S_B 
  + S_\psi + 2Ha^2_\pm S^\pm_\xi
\right) \,,
\nonumber\\
\end{eqnarray}
where the radial gauge of the spatial component at $J$th order is the isotropic gauge, as before.

\subsection{ Evaluation of perturbations}

Now we can explicitly evaluate the gravity induced on the branes. To simplify the analysis, we assume that matter fields exist on one of the two branes, which is regarded as the visible brane. With this simplification, the sum of the Newton potentials $\Phi_\pm ~(:=4 \pi G \Delta^{-1} \rho_\pm^{(1)})$ which generally appears in the two-brane system as we see in Eq. (\ref{eq:Y^J_[GGS] y_+-}) is replaced as $\sum_{\sigma=\pm} \Phi_\sigma \to \Phi_\pm$ at $y=y_\pm$.

First, we discuss linear perturbations. 
Because the evaluation of the spatial component is almost the same as that of the temporal component, we henceforth concentrate on the temporal component. Assuming weak back-reaction  $a\approx e^{-|y|/\ell}$, the results are summarized as 
\begin{eqnarray}
 \bar A^{(1)}_{+} (r,y_{+})
&=& 
 2\Phi_+ 
+ O\left( r_\star^2 \Delta \Phi_+  \right)  
\biggl\{
   O\left( \beta_+  \right) 
 + O\left( \gamma_+ \right) 
 - \beta_+  
     \Bigl[  \Bigl(\frac{\dot H_+}{3H_+^2} \Bigr)^2  \alpha_+ 
     +    \alpha_-  \Bigr]
\biggr\},
\nonumber
\\
 \bar A^{(1)}_{-} (r,y_{-})
&=& 
  2 \Phi_-  
+ O\left( r_\star^2 \Delta \Phi_+  \right)  
\biggl\{
   O\left( \beta_-  \right) 
 + O\left( \gamma_- \right) 
 - \beta_-  \bigl[ a^4_-  \alpha_+ + \alpha_- \bigr]
\biggr\} \,, 
\qquad   
 \label{eq:sol bar A^2_[000]}
\end{eqnarray}
where $\alpha_\pm$, $\beta_\pm$ and $\gamma_\pm$ are defined as   
\begin{eqnarray}
 \alpha_\pm &:=&  \frac{2N^3 \lambda_\pm}{ \kappa \dot\phi_{0}^2(y_\pm)} 
             = - \frac{2N^3 \lambda_\pm}{ 3 \dot H_\pm } \,, 
\nonumber
\\
 \beta_\pm
   &:=&   \frac{\ell^2}{a^4_\pm  r_\star^2} 
    = \left\{
\begin{minipage}[c]{8cm}
$\displaystyle
   \hspace{50pt}  \frac{\ell^2}{r_\star^2},  
   \hspace{50pt} (y=y_+) \,,$
\newline
$\displaystyle
    \Bigl( \frac{ 0.1 {\mathrm{mm}}}{r_\star} \Bigr)^2
    \Bigl( \frac{ 10^{-16} }{a_-}  \Bigr)^4
    \Bigl( \frac{ \ell}{\ell_{Pl}}  \Bigr)^2,
     \quad ( y=y_-),$
\end{minipage} 
\right.
\label{eq:suppres factor 1}
\\
\gamma_\pm &:=&
     (a_{\pm}^4 \tilde m_S^2 r_\star^2)^{-1}
     = \beta_\pm ( \tilde m_S \ell )^{-2} \,.
\nonumber
\end{eqnarray}
Here, $\alpha_\pm$ is the correction from the interaction terms, the massive mode contributions of the TT part are expressed by $\beta_\pm$, and the short-range contributions of the scalar-type perturbations are represented by $\gamma_\pm$, in which we have used the effective radion mass $\tilde{m}_S$. When the radius stabilization mechanism proposed by Goldberger and Wise is effective, the mass $\tilde m_S$ becomes $O(\ell^{-1})$.\cite{Tanaka:2000er,Goldberger:1999uk}

Let us first consider the corrections induced on the negative tension brane ($y=y_-$).  Because we are concerned with the hierarchy resolution in this case,\cite{Randall:1999ee} we have set the AdS curvature length $\ell$ and the hierarchy $a_+/a_-$ to the Planck length $\ell_{Pl}$ and $10^{16}$, respectively. 
Then the expression for $\beta_-$ implies the dominant contribution of the massive mode for length scales $\lesssim 0.1$mm. However the KK mode does not contribute to the force outside the matter distribution but, rather, to the matter energy density, since the gravitational potential appears only in the form $\Delta \Phi_\pm$, which is proportional to $\rho_\pm$.
Hence the change of the metric perturbation due to the short-range part becomes significant on the negative tension brane only when $\rho^{(1)}_- \gtrsim O(\mathrm{TeV}^4)$. This is also true for the correction from $\gamma_-$ as long as the effective radion mass is $O(\ell^{-1})$.
For $\alpha_\pm$, the evaluation depends on the details of the stabilization model. This correction becomes important compared to that from the short-range part when $a^4_- \alpha_+$ or $\alpha_-$ exceeds $O(1/\tilde{m}_S^2 \ell^2)$.
We note that when $a_-^4 \alpha_+\gg 1$, the mass of the stabilized radion becomes small, and then the effect appears also through $\gamma_-$, although this case seems to be exceptional.\cite{Kudoh:2002mn}

On the positive tension brane, the corrections from $\beta_+$ and $\gamma_+$ are similar to those in the one-brane model, and hence are suppressed. 
As we pointed out, the difference is that the corrections in this case are proportional to the matter energy density.
The factor of $\alpha_+$ is suppressed only by $\dot H_+/H_+^2$, which is small but not hierarchically suppressed. 
As long as $\lambda_+$ has a natural order of magnitude smaller than
$\ell$, $\alpha_+$ is at most $O(H_+^2/\dot H_+)$. 
Then, the correction remains less than 
$O(\beta_{+})O(r_\star^2 \Delta \Phi_+)$.
However, when $\lambda_+$ is much larger than $\ell$, the correction becomes larger than that in the TT part by a factor of $\lambda_+/\ell$. 
Although these choices of parameters are not natural, the possibility of an enhanced correction might be interesting.

Let us discuss second-order perturbations. 
After straightforward but tedious calculations, it is shown on each brane that the leading order in the derivative expansion is identical to 4D Einstein gravity even in second-order perturbations. 
Thus to complete the recovery, it is necessary to investigate the contributions from the massive modes and the interaction terms, which appear at the next order in the derivative expansion. However, we immediately realize that at this order there are some terms on the negative tension brane that are hierarchically enhanced by a factor of $O(\beta_-/a_-^{2})$ in comparison to the usual post-Newtonian terms. It therefore seems that the recovery of the conventional 4D gravity is spoiled at this order.
This point is studied in Refs. \citen{Kudoh:2001kz} and \citen{Kudoh:2002mn}. 
As discussed there, however, these dangerous terms are completely canceled out. The remaining contributions are actually suppressed apart from some exceptional cases which come from $\alpha_\pm$, as mentioned above.

The results are summarized as 
\begin{eqnarray}
\Delta \bar A^{(2)}_{+} (r,y_{+})
&=&  
   8 \pi G 
   (\rho^{(2)}_{+} + 3 P^{(2)}_{+})
   - 4 \Phi_+  {\Delta}  \Phi_+ 
\nonumber
\\
&&
+ O\left( \frac{\Phi_+^2}{r_\star^2} \right)  
\biggl\{
   O\left( \frac{\beta_+}{a_-^2}  \right) 
 + O\left( \frac{\gamma_+}{a_-^2}   \right) \nonumber\\
&&\hspace{2cm}
 + O\left( \beta_+ \right)
     \Bigl[  \Bigl(\frac{\dot H_+}{ H_+^2} \Bigr)^2  O( \alpha_+ )
     +  O \Bigl( \frac{ \alpha_- }{a^2_-} \Bigr) \Bigr]
\biggr\},
\nonumber
\\
\Delta \bar A^{(2)}_{-} (r,y_{-})
&=& 
   8 \pi G 
   (\rho^{(2)}_{-} + 3 P^{(2)}_{-})
   - 4 \Phi_-  {\Delta}  \Phi_- 
\nonumber
\\
&&
+ O\left( \frac{\Phi_-^2}{r_\star^2} \right)  
\biggl\{
   O\left( \beta_-  \right) 
 + O\left( \gamma_- \right) 
 + O\left( \beta_- \right)
     \bigl[ O(a^4_-  \alpha_+) + O(\alpha_-)  \bigr]
\biggr\}.\nonumber\\
\label{eq:sol bar A^2_[000]}
\end{eqnarray}
In the case that matter fields are confined to the positive tension brane, the corrections from KK modes are multiplied by a factor of $O(\beta_+/a_-^{2})$ compared to the usual post-Newtonian terms.  
Nevertheless, the appearance of an enhancement by a factor of $1/a_-^2$ is very likely an artifact of the derivative expansion method, although the corrections are still suppressed.  Since the condition that the typical length scale of the spatial gradient is larger than that of the change in the fifth direction becomes $(\ell^2/a^2_- r_\star^2) = (\beta_+/a^2_-)\ll 1$ near the negative tension brane, $\beta_+/a^2_-$ appears as an expansion parameter. Although the correction seemingly becomes large in the $y_- \to \infty$ limit, this is due to the limitation of the present approximation.

\subsection{Convention for the length scale}
\label{sec:convention}

To end this section, we comment on the convention used for the length scale that we have adopted in the previous sections.
In our study of the two-brane model, we have used the normalization scheme in which the length scale is always measured on the positive tension brane at $y=y_+$. Thus, when we regard the negative tension brane at $y=y_-$ as the visible brane and investigate physical quantities on this brane, we must rescale them to obtain their values on the hidden brane at $y=y_+$. This is the reason that we have used the definition of the energy-momentum tensor (\ref{eq:EM tensor+-}). It is sometime convenient to measure a length scale $r_\star$ with respect to its proper length scale $a(y) r_\star$ at the location $y$. This scheme is adopted in Ref. \citen{Tanaka:2000er}.

To use the proper length normalization scheme in our analysis, we must change some notation and definitions. First, the energy-momentum tensors are defined as 
\begin{eqnarray}
    T_{{\pm} \nu }^{ ~~ \mu} = {\mathrm{diag}} 
    \{- \rho_{\pm}, P_{\pm}, P_{\pm}, P_{\pm} \},
\label{eq:EM tensor:N2}
\end{eqnarray}
where $T_{{+} \nu }^{ ~ \mu}$ and $T_{{-} \nu }^{~ \mu}$ are physical quantities measured on the positive and the negative tension branes, respectively.
The induced 4D Newton's constant and the Newton potential are defined on each brane as 
\begin{eqnarray}
      8\pi G_{\pm} &:=& \kappa N a_{\pm}^2 \,,
\label{eq:4D Newton cons:N2}
\\
  \frac{1}{a_{\pm}^2} \Delta \Phi_\pm (r) 
    &:=&
      4 \pi G_\pm \rho^{(1)}_{\pm}   \,.
\label{eq:def Newton potential:N2}
\end{eqnarray}
The Laplacian operator must accompany the warp factor, i.e., $a^{-2}_\pm\Delta$.  In spite of these changes, the results of the linear perturbations are the same, $ \bar A^{(1)}_0 =2  \Phi_\pm$. 
The leading terms of the second-order perturbations (\ref{eq:sol bar A^2_[000]}) are given in this normalization as 
\begin{eqnarray}
   \frac{\Delta}{a^2_\pm} \bar A^{(2)} (r,y_{\pm})
   &=&  8 \pi G_{\pm}
      (\rho ^{(2)}_{\pm} + 3P ^{(2)}_{\pm}  )
 - 4 \Phi_\pm \Bigl(\frac{\Delta}{a_{\pm}^2} \Bigr)
     \Phi_\pm
\,.
\end{eqnarray}
The modification just consists of the extra warp factor for the Laplacian and the 4D Newton's constant. Therefore the analyses presented in previous sections can be applied here too with only slight changes.

However, it is necessary to modify the interpretation of gravitational coupling scale.
Recall that when we set $y=y_-$ ($y=y_+$), the negative (positive) tension brane is regarded of as the visible brane, and we ignore the matter fields on the other, hidden brane.
When the negative tension brane is the visible brane, the hierarchy resolution is involved, and then the fundamental scale is set to the TeV scale, as in the original RS two-brane model. 
On the other hand, when the positive tension brane is regarded as the observable brane, we take the fundamental scale as the Planck scale, as in the RS one-brane model. 
Because the gravitational coupling scale on each brane is warped, as described by Eq. (\ref{eq:4D Newton cons:N2}), we must vary the value of $\kappa N$ according to which brane is visible. 
 To obtain the corresponding gravitational coupling scale at $y=y_+$ or $y=y_-$, the modification is given as 
\begin{eqnarray}
\kappa N 
\to 
 \left\{
\begin{array}{ccc}
\displaystyle
   O( M_{\rm Pl}^{-2})  &\quad ( y=y_+),&
\cr
\cr
\displaystyle
   O( M_{\rm EW}^{-2})  &\quad ( y=y_-),&
\end{array} 
\right.
\label{eq:def. fundamental scale}
\end{eqnarray}
where $M_{\rm Pl}$ is the 4D Planck mass and $M_{\rm EW}$ is a mass scale of order TeV. 
With this modification, we must change the suppression factors. They are given by 
\begin{eqnarray}
 \beta_\pm
    = \left\{
\begin{array}{ccc}
\displaystyle
    \frac{\ell_{\rm Pl}^2}{r_\star^2}, 
   &\quad ( y=y_+) & 
\cr
\displaystyle
    \Bigl( \frac{0.1 mm}{a_- r_\star} \Bigr)^2
    \Bigl( \frac{10^{-16}}{a_-} \Bigr)^2
    \Bigl( \frac{ \ell_{\rm EW}}{10^{-2} \mathrm{TeV}^{-1} }  \Bigr)^2,
    &\quad (y=y_-)&
\end{array} 
\right.
\end{eqnarray}
where  $\ell_{\rm Pl} \approx M_{\rm Pl}^{-1}$ and $\ell_{\rm EW} \approx
M_{\rm EW}^{-1}$. 
With these modifications, the analysis given in the previous sections for the recovery of 4D Einstein gravity is valid.

    \section{Summary}
    \label{sec:summary}

We have considered second-order gravitational perturbations in the RS one-brane model and in the RS two-brane model with the radius stabilization mechanism. It is shown in each model that second-order perturbations behave well, and the results are basically consistent with 4D Einstein gravity, whose temporal component at second order and spatial component at first order have been observationally confirmed to an accuracy of about 0.1 {\%}.\cite{will} 

In the RS one-brane model, deviations appear as a change of Newton's law and are suppressed by a factor of $O( (\ell^2/r_\star^2) \ln (r_\star/\ell))$. Thus it is not feasible to observationally distinguish these corrections.

For the RS two-brane model, the analysis becomes slightly complicated since, owing to the stabilized radius, there are scalar-type perturbations in addition to the transverse-traceless perturbations. 
As a model of radius stabilization, we have assumed a scalar field that has a potential in the bulk and a potential on each brane. 
 Three types of corrections are found:  massive mode contributions of transverse-traceless perturbations, massive mode contributions of scalar-type perturbations, and corrections from the interaction terms on the brane.

When we consider the case in which the matter fields are on the negative tension brane, the correction to 4D Einstein gravity appears at a relative order of $O((a_+/a_-)^{4}(\ell/r_\star)^2)$.
With the choice of the hierarchy (\ref{eq:suppres factor 1}), the correction to the metric in the linear perturbations becomes comparable to the usual Newtonian potential when 
$r_\star \lesssim 0.1$mm. However, this correction does not give a contribution to the force outside the matter distribution. 
Hence, it seems that this correction dose not prevent reproductions of the predictions of 4D Einstein gravity. 
We have not confirmed if this feature remains in second-order perturbations. Nevertheless, the corrections are suppressed by the above factor compared to the usual post-Newtonian terms, and therefore the effect due to these corrections is almost impossible to detect by observing (post-) Newton's law.

In the case that the matter fields are on the positive tension brane, the correction to 4D Einstein gravity given by linear-order perturbations appears at a relative order of $O((\ell/r_\star)^2)$. 
The corrections given by second-order perturbations are 
$O((a_+/a_-)^2 (\ell/r_\star)^2)$ compared to the usual post-Newtonian terms. 
Although the corrections are still small, the unexpected extra factor of $(a_+/a_-)^2$ might suggest that deviations from 4D Einstein gravity appear in higher-order perturbations. 
However, this is very likely to be an artifact due to the limitation of our approximation scheme.

To give a complete proof of the recovery of 4D Einstein gravity, further extension of the present analysis is necessary, and a clearer understanding of the recovery mechanism in the regime beyond linear perturbations is also necessary.

\section*{Acknowledgements}
The author would like to thank Takahiro Tanaka for reading of the manuscript.  The author is supported by JSPS Research Fellowships for Young Scientists.



\begin{thebibliography}{99}


\bibitem{Arkani-Hamed:1998rs}
N.~Arkani-Hamed, S.~Dimopoulos and G.~R.~Dvali,
Phys.\ Lett.\ B {\bf 429}  (1998), 263.\\
 I.~Antoniadis, N.~Arkani-Hamed, S.~Dimopoulos and G.~R.~Dvali,
Phys.\ Lett.\ B {\bf 436} (1998), 257.

 
\bibitem{Randall:1999ee}
L.~Randall and R.~Sundrum,
Phys.\ Rev.\ Lett.\  {\bf 83} (1999), 3370.
 

\bibitem{Randall:1999vf}
L.~Randall and R.~Sundrum,
Phys.\ Rev.\ Lett.\  {\bf 83} (1999), 4690.

\bibitem{Shiromizu:2000wj}
T.~Shiromizu, K. ~Maeda and M.~Sasaki,
Phys.\ Rev.\ D {\bf 62} (2000), 024012.

\bibitem{Sasaki:2000mi}
M.~Sasaki, T.~Shiromizu and K.~Maeda,
Phys.\ Rev.\ D {\bf 62}  (2000), 024008.
 
 
\bibitem{Garriga:2000yh}
J.~Garriga and T.~Tanaka,
Phys.\ Rev.\ Lett.\  {\bf 84} (2000), 2778.

\bibitem{Tanaka:2000er}
T.~Tanaka and X.~Montes,
Nucl.\ Phys.\ B {\bf 582} (2000), 259.

 
\bibitem{Mukohyama:2001ks}
S.~Mukohyama and L.~Kofman,
Phys. Rev. D {\bf 65} (2002), 124025.
 
 
\bibitem{Mukohyama:2001jv}
S.~Mukohyama,
Phys. Rev. D {\bf 65}  (2002),  084036.

\bibitem{Tanaka:2000zv}
T.~Tanaka,
Prog.\ Theor.\ Phys.\  {\bf 104} (2000), 545.

\bibitem{Giddings:2000mu}
S.~B.~Giddings, E.~Katz and L.~Randall,
J. High Energy Phys.  {\bf 03} (2000), 023.

 

 
\bibitem{Wiseman:2001xt}
T.~Wiseman,
Phys. Rev. D {\bf 65} (2002), 124007.
.
 
 
\bibitem{Giannakis:2001zx}
I.~Giannakis and H.~C.~Ren,
Phys.\ Rev.\ D {\bf 63} (2001), 024001.


\bibitem{Kudoh:2001wb}
H.~Kudoh and T.~Tanaka,
Phys.\ Rev.\ D {\bf 64}  (2001), 084022.
 

\bibitem{Kudoh:2001kz}
 H.~Kudoh and T.~Tanaka,
Phys. Rev. D {\bf 65} (2002), 104034.  
 

\bibitem{Kudoh:2002mn}
H.~Kudoh and T.~Tanaka,
hep-th/0205041.
 


\bibitem{Goldberger:1999uk}
W.~D.~Goldberger and M.~B.~Wise,
Phys.\ Rev.\ Lett.\  {\bf 83}  (1999),  4922.

    
\bibitem{will} 
    C.M. Will,  \textit{Theory and Experiment in Gravitational Physics} revised ed. (Cambridge University Press, Cambridge, England, 1993).

\end{thebibliography}
\end{document}